\documentclass[3p]{elsarticle}

\usepackage{a4wide}
\usepackage{float}
\usepackage{amsmath}
\usepackage{amssymb}
\usepackage{natbib}
\usepackage{graphicx}
\usepackage{caption}
\usepackage{subcaption}
\usepackage[latin1]{inputenc} %Kann auch Umlaute etc. erkennen
\usepackage{rotating}
\usepackage{setspace}
\usepackage{sidecap}
\usepackage{upgreek}
\usepackage{microtype}
\usepackage{url}
\usepackage{lineno}
\usepackage{multirow}
\usepackage{booktabs}
 \biboptions{sort&compress}

\linenumbers
\begin{document}
\begin{frontmatter}

%\pagenumbering{Roman}
% avoid new pages for individual chapters
\let\clearpage\relax
%\include{chapters/title}
%------------------------------------ Title ------------------------
\title{Determination of the electronics transfer function for current transient measurements}

\renewcommand{\thefootnote}{\fnsymbol{footnote}}

\author[]{Christian Scharf\corref{cor1}}
\author[]{Robert Klanner}
%\author[]{Robert Klanner}

\cortext[cor1]{Corresponding author. Email address: Christian.Scharf@desy.de, Telephone: +49 40 8998 4725.}
\address{~Institute for Experimental Physics, University of Hamburg, Germany}
%\address{~Institute for Experimental Physics, University of Hamburg, Luruper Chaussee 149, 22761 Hamburg, Germany}

%\onehalfspacing
\begin{abstract}

 We describe a straight-forward method for determining the transfer function of the readout of a sensor for the situation in which the current transient of the sensor can be precisely simulated.
 The method relies on the convolution theorem of Fourier transforms.
 The specific example is a planar silicon pad diode.
 The charge carriers in the sensor are produced by picosecond lasers with light of wavelengths of 675 and 1060\,nm.
 The transfer function is determined from the 1060\,nm data with the pad diode biased at 1000\,V.
 It is shown that the simulated sensor response convoluted with this transfer function provides an excellent description of the measured transients for laser light of both wavelengths.
 The method has been applied successfully for the simulation of current transients of several different silicon pad diodes.
 It can also be applied for the analysis of transient-current measurements of radiation-damaged solid state sensors, as long as sensors properties, like high-frequency capacitance, are not too different.
\end{abstract}

\end{frontmatter}
%
%\tableofcontents

%\newpage

%\section{Introduction}
%\label{sect:Introduction}

 \paragraph{Introduction}
 The analysis of current transients from different sensors is frequently limited by the knowledge of the electronics response, which is also influenced by the sensor properties.
  Examples for silicon sensors are the determination of the electric fields, carrier lifetimes and charge multiplication in radiation-damaged sensors using the Transient Current Technique (TCT, edge-TCT)\,~\cite{Kramberger:2010, Mikuz:2011, Pacifico:2011} or charged particles with shallow incident angles\,~\cite{Swartz:2006}.
 The aim of this Technical Note is to demonstrate that, for an experiment in which the pulse shape from the sensor can be precisely simulated, the electronics transfer function can be obtained from the measured transient using the convolution theorem of Fourier transforms.
 This transfer function can then be used for analyzing data for which the pulse shape of the sensor is not known.
 An example is the analysis of measured transients from a radiation-damaged sensor using the known transfer function obtained from the sensor before irradiation, as long as detector properties, like the high-frequency capacitance, do not change too much with irradiation.

 \paragraph{Measurement set-up}
 The measurement set-up used is described in detail in\,~\cite{Scharf1:2014, Becker:2011, Becker:2010}.
 The measurements have been performed on $p^+n\,n^+$ and $n^+p\,p^+$ pad diodes with different doping, thicknesses of $200\,\upmu$m and $285\,\upmu$m, and 4.4\,mm$^2$ and 25\,mm$^2$ area. In all cases the electronics response function has been successfully determined. Here we present the results from a $p^+n\,n^+$ pad diode produced by Hamamatsu\,~\cite{Hamamatsu} on a $\langle 100 \rangle $ crystal with 204.5\,$\upmu $m mechanical thickness, 4.4\,mm$^2$ area, and $ 2.9 \cdot 10^{12}$\,cm$^{-3}$ phosphorous doping, which was connected by a 3\,m long RG58 cable and a bias-T to an amplifier\,~\cite{Ampli}, and read out by a Tektronix DPO 4104 oscilloscope with a bandwidth of 1 GHz and a sampling rate of 5 GS/s. A guard ring was present but not connected for the measurements.
 The active thickness of the pad diode was estimated to $200 \pm 1\,\upmu $m from dielectric as well as caliper measurements of the physical thickness minus the thickness of the implants from spreading resistance measurements\,~\cite{HEPHY:2013}. The depletion voltage was determined to $87.5 \pm 3.0$\,V, from capacitance measurements with a capacitance above the depletion voltage of about $2.7$\,pF.

 The charge carriers were generated by picosecond lasers\,~\cite{Laser} pulsed at a frequency of 200\,Hz with a full-width-at-half-maximum of less than 50\,ps and wavelengths of 675 and 1060\,nm.
 For each pulse approximately $10^6$\,electron-hole\,pairs were generated, and for every measurement 512 pulses were averaged.
 At room temperature the absorption length in silicon for light of 1060\,nm is about 1.5\,mm.
 As the attenuation length is long compared to the sensor thickness, the distribution of charge carriers is similar as for charged particles traversing the sensor.
 At this wavelength the absorption length is a strong function of temperature\,~\cite{Sze:2006, Kramberger:PhD}.
 It is about $650\,\upmu $m at $40^\circ $C.
 The absorption length for light of 675\,nm is about 3.3\,$\upmu $m at room temperature, and the signal induced in the electrodes of the sensor is essentially due to electrons if the light is injected at the $p^+$ side, and due to holes if injected at the $n^+$ side.

 \paragraph{Simulations}
 In the simulations a uniform doping in the active region of the sensor is assumed, resulting in a linear position dependence of the electric field.
 Using the field simulated with SYNOPSYS-TCAD\,~\cite{TCAD} which includes realistic doping distributions at the $p^+n$ and $n^+n$ transitions\,~\cite{JSchwandt}, it has been checked that the current transients for voltages 50\,V above the depletion voltage are hardly affected by the electric field distribution at the transitions.

 Electrons and holes are generated on a grid with 100\,nm spacing according to exponentials with the light-absorption lengths given above.
 The charge carriers are then drifted in the electric field  in time steps  $\Delta t =10$\,ps taking into account diffusion by Gaussians with variances $\sigma_{e} = \sqrt{(2\cdot \mu_{e} \cdot k_B T/q_0) \Delta t}$ for electrons, and similar for holes, with the Boltzmann constant $k_B$, the absolute temperature $T$, the elementary charge $q_0$, and the electron and hole mobilities $\mu _e$ and $\mu _h$.
 The field dependence of the mobilities of electrons and holes was adjusted to describe our measurements\,~\cite{Scharf:2014, Scharf1:2014}. The main difference compared to the standard parametrization\,~\cite{Jacoboni:1977} is that at high fields ($\approx50\,$kV/cm) the electron and the hole drift velocities are similar. We note that there are hardly any measurements of the drift velocities for $\langle 100 \rangle $ silicon available.
 The current induced in the electrodes in the time interval between $i \cdot \Delta t$ and $(i+1) \cdot \Delta t$ is calculated according to $I_{i}^{sim} = q_0 /\Delta t \cdot \sum_j{\big[(N_{i+1,j}^e - N_{i,j}^e) - \big(N_{i+1,j}^h - N_{i,j}^h\big)\big]}$, where $N_{i,j}^{e}$ is the number of electrons and $N_{i,j}^{h}$ the number of holes at the grid point $j$ at time $i \cdot \Delta t$.
 Effects like charge trapping or charge multiplication are not taken into account.
 The convoluted signal is given by $ S_{k}^{sim} = \sum_l I_{k-l}^{sim} \cdot R_l$, where $R_l$ is the response at $t=l \cdot \Delta t$ to an initial unit current at $t = 0$.

 \paragraph{Transfer function determination}
 The measurements have been performed for voltages between 100\,V and 1000\,V in steps of 10\,V.
 For every voltage three sets of data were taken: 1060\, nm light injected from the $p^+$ side, called \emph{"IR"}, 675\,nm light injected from the $p^+$ side, \emph{"e"}, and 675\,nm light injected from the $n^+$ side, \emph{"h"}.
 For \emph{"IR"} holes and electrons contribute equally to the signal, whereas for \emph{"e"} electrons, and for \emph{"h"} holes dominate.
 For determining the transfer function $R$, the \emph{IR} measurement at 1000\,V is used.
 A spline interpolation of the measurement $I^{int}$ is used to obtain values for the same time steps $\Delta t = 10$\,ps as in the simulation.
 Next the Fast Fourier Transforms $\cal F $($I^{sim})$ and $\cal F $($I^{int})$ are calculated, and the transfer function is obtained by
 $R = \cal{F}$\,$^{-1}\big[\cal F$($I^{int})/\cal F$ ($I^{sim})\big]$, using the well-known convolution theorem $\cal F$($f \otimes g) = $ $\cal F$($f) \cdot \cal F$($g$).

 \begin{figure}[!ht]
 \centering
  \includegraphics[width=11.5cm]{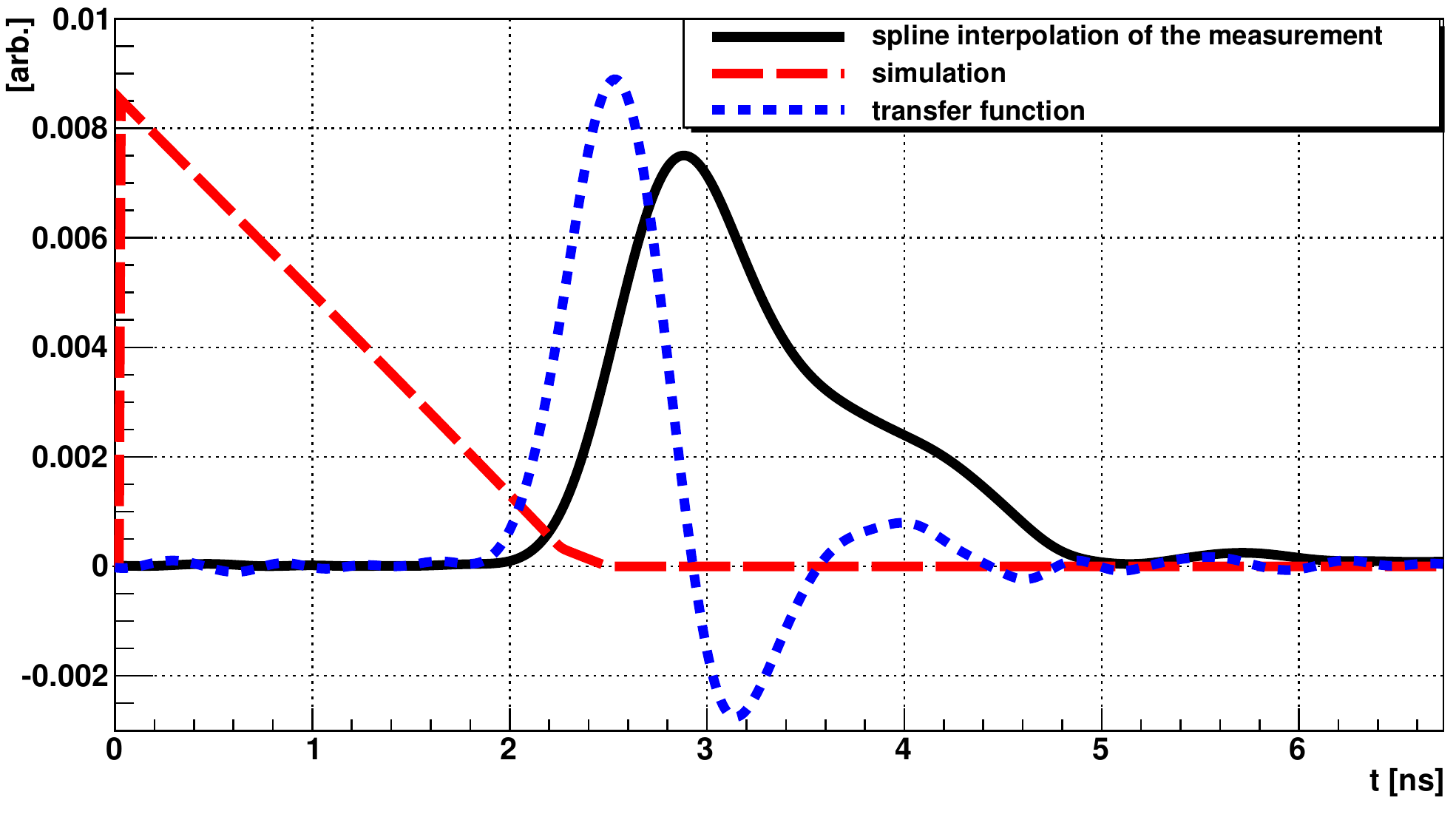}
   \caption{Determination of the transfer function $R$ (blue dots) from the spline interpolation of the measured transient $I^{int}$ (black solid) and the simulated transient $I^{sim}$ (red dashed) for the signal of electrons and holes produced by 1060\,nm laser light (\emph{IR}) at 1000\,V.
   }
 \label{fig:Fig-Conv}
\end{figure}

 Fig.\,\ref{fig:Fig-Conv} shows the elements used in the determination of the transfer function $R$: the spline-interpolated measured current transient $I^{int}$, the simulated current transient before convolution $I^{sim}$, and $R$ obtained as described. As shown in Fig.\,\ref{fig:Fig-EH1000} the transients for electrons and holes at 1000\,V are very similar implying that at high fields around 50\,kV/cm the drift velocities of electrons and holes are also similar. As a result, $I^{sim}$ for \emph{IR} at 1000\,V, which is a superposition of electron and hole drift, is to a good approximation a straight line without a tail from slower holes.
 The time step for all curves is 10\,ps. The time shift between the simulated and the measured transient is arbitrary.
   It does not change the shape of R, but just its position along the time axis.

  \begin{figure}
%  [!ht]
   \centering
   \includegraphics[width=8.0cm]{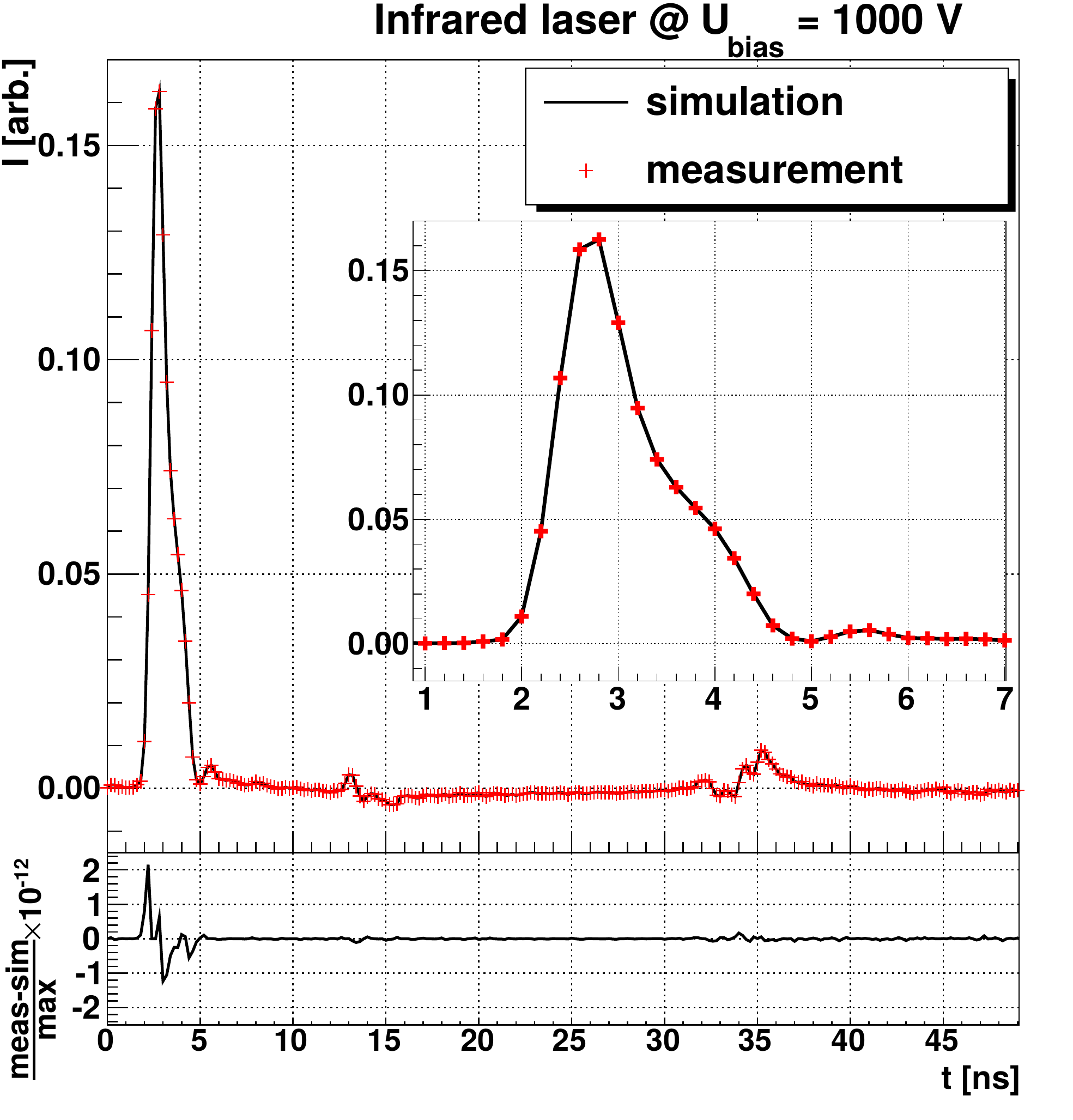}
    \caption{Comparison of the measured current transient (crosses) with the simulated one (solid line) for \emph{IR} at 1000\,V. The transfer function used for the convolution has been determined from this measurement.}
  \label{fig:Fig-IR1000}
 \end{figure}

 \paragraph{Comparison of measurements with simulations}
 Fig.\,\ref{fig:Fig-IR1000} compares the current transient for the \emph{IR} measurement at 1000\,V with the simulated transient using for the convolution the transfer function determined from the same data. For the simulation only the values at the times at which data were recorded are shown. Both the main pulse, shown as insert, as well as the signal reflections are well described.
 The difference between the measured and the simulated signal divided by the maximum value of the measured signal is shown at the bottom. Its absolute value is smaller than $10^{-11}$.
 This demonstrates the consistency of the method used for determining the electronics transfer function.

   \begin{figure}[!ht]
   \centering
%   \begin{subfigure}[a]{0.5\textwidth}
%    \includegraphics[width=\textwidth]{Setup.pdf}
   \begin{subfigure}[a]{7.5cm}
    \includegraphics[width=7.5cm]{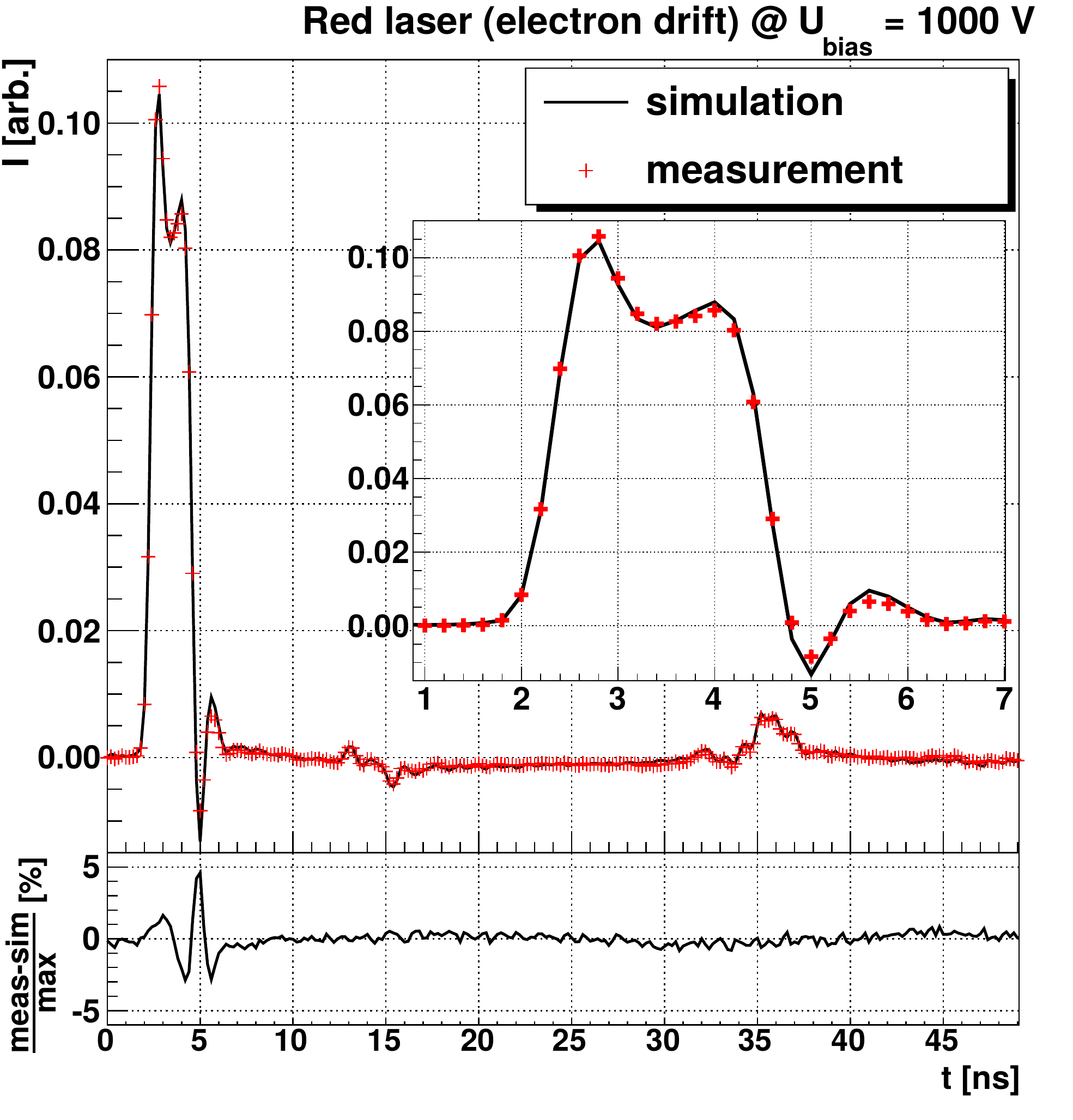}
    \caption{ }
     \label{fig:E1000}
   \end{subfigure}%
    ~
%   \begin{subfigure}[a]{0.5\textwidth}
%    \includegraphics[width=\textwidth]{Source.pdf}
   \begin{subfigure}[a]{7.5cm}
    \includegraphics[width=7.5cm]{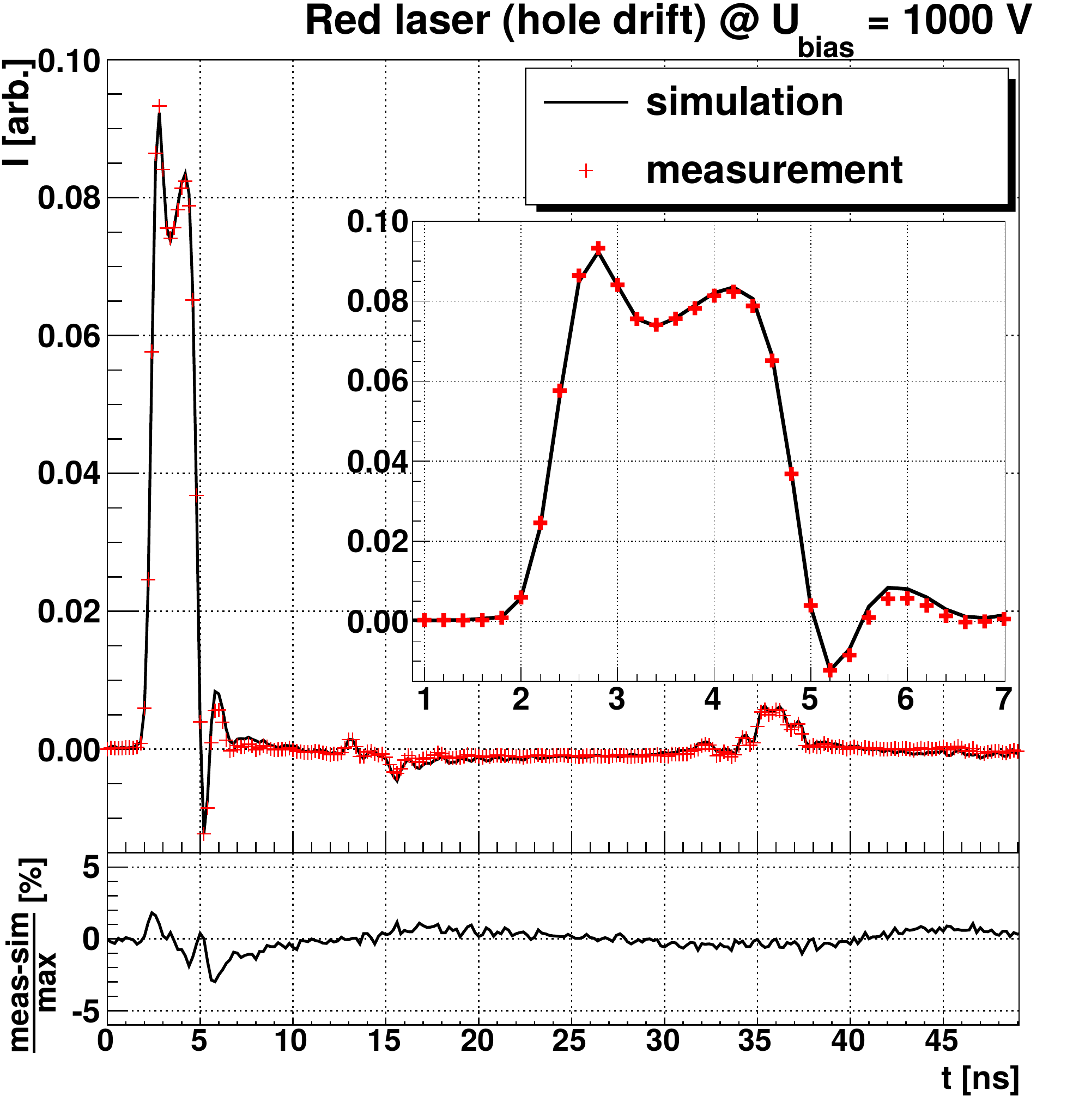}
    \caption{ }
     \label{fig:H1000}
   \end{subfigure}%
   \caption{\,Comparison of the measured current transients (crosses) with the simulated ones (solid lines) at 1000\,V using the same transfer function for the convolution as in Fig.\,\ref{fig:Fig-IR1000} for (a)\,electrons, and (b)\, holes.}
  \label{fig:Fig-EH1000}
\end{figure}

 \begin{figure}[!ht]
 \centering
 \includegraphics[width=11.0cm]{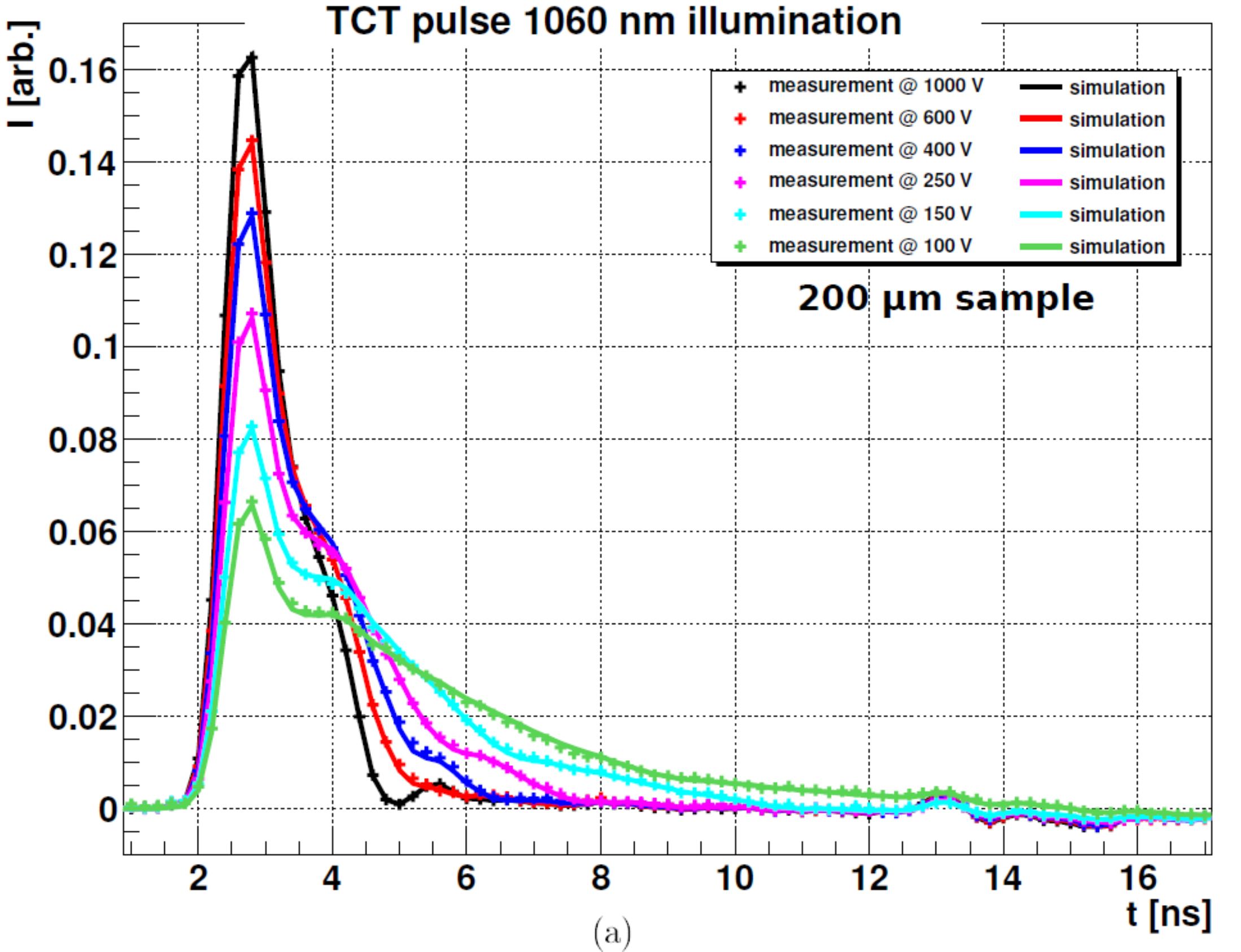}
 \caption{Current transients measured (crosses) and simulated (solid lines) for different bias voltages for 1060\,nm laser light.}
 \label{fig:Fig-IRDrift}
\end{figure}

 \begin{figure}[!ht]
 \centering
 \includegraphics[width=11.0cm]{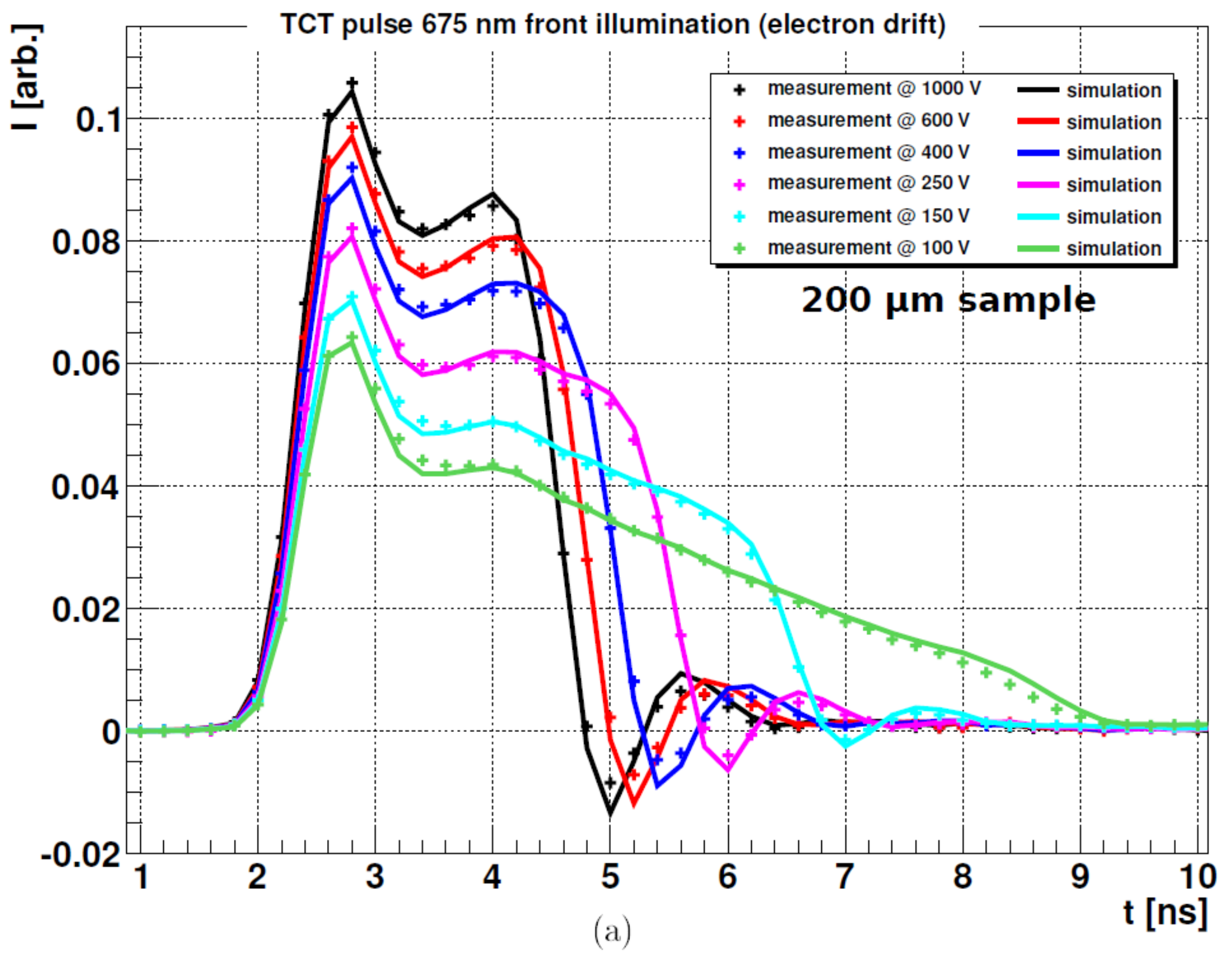}
 \caption{Current transients measured (crosses) and simulated (solid lines) for different bias voltages for 675\,nm laser light injected from the $p^+$ side.}
 \label{fig:Fig-EDrift}
\end{figure}

 \begin{figure}[!ht]
 \centering
 \includegraphics[width=11.0cm]{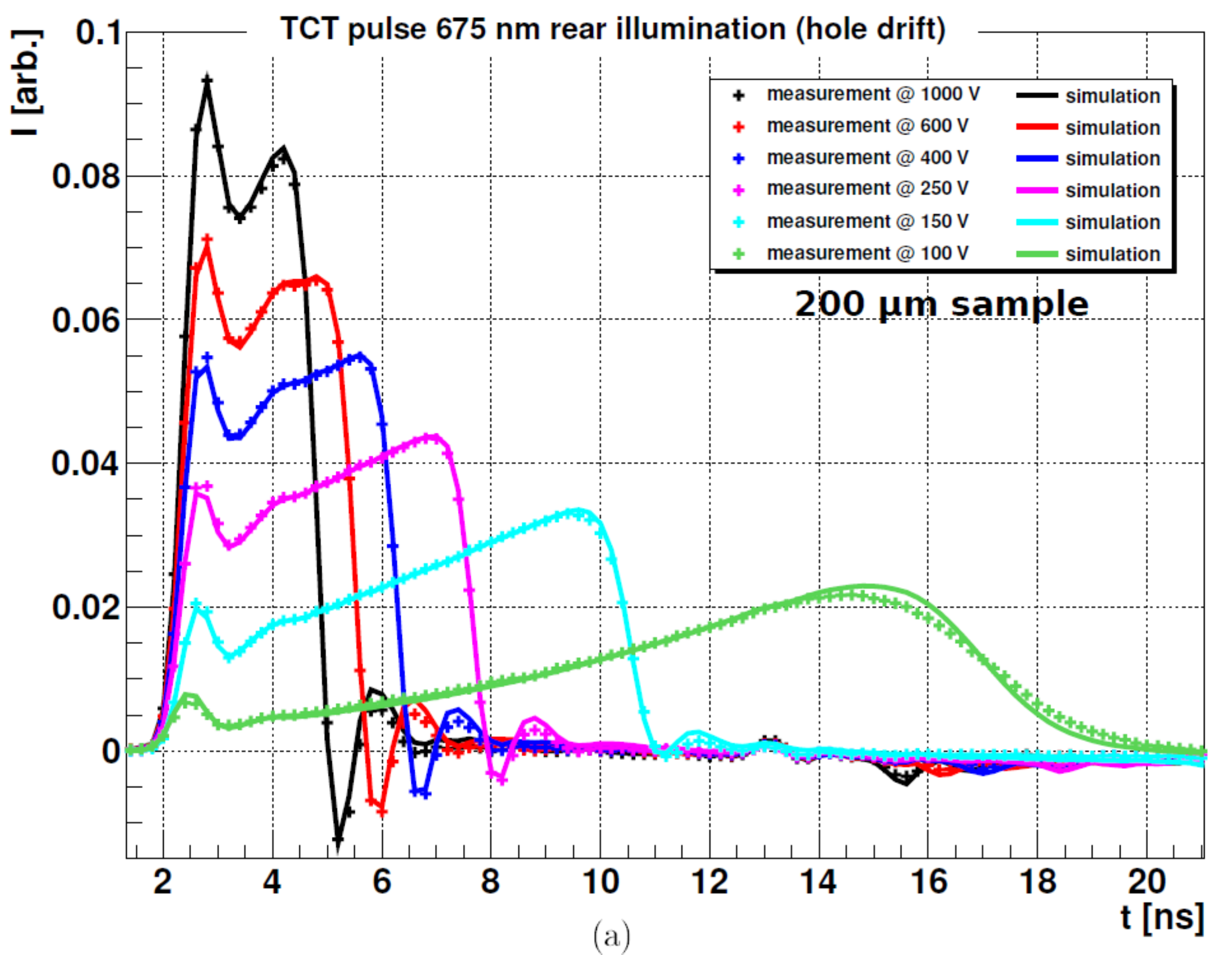}
 \caption{Current transients measured (crosses) and simulated (solid lines) for different bias voltages for 675\,nm laser light injected from the $n^+$ side.}
 \label{fig:Fig-HDrift}
\end{figure}

  Fig.\,\ref{fig:Fig-EH1000} compares the current transients for the $e$ and for the $h$ measurement at 1000\,V with the simulated transients using for the convolution the transfer function determined from the \emph{IR} measurement at 1000\,V.
  We note that the transients are very similar and that the $h$ pulse is only slightly longer than the $e$ pulse. At 1000\,V the electric field in the sensors varies between 46\,kV/cm and 54\,kV/cm. We conclude that for $\langle 100 \rangle $ silicon the drift velocities of electrons and holes at these high fields are similar with little dependence on the electric field. 
  The difference between the measured and the simulated signal divided by the maximum value of the measured signal is displayed at the bottom. Its absolute value is less than 5\,\%.
  This difference provides an idea of the accuracy of the simulation and the method used for determining the transfer function since the simulated transient before convolution for \emph{IR} is significantly different compared to the ones for \emph{e} and \emph{h} which are to first approximation boxcar functions.

 Fig.\,\ref{fig:Fig-IRDrift} compares the measured with the simulated current transients for bias voltages between 100 and 1000\,V for $IR$.
 It is seen that for all voltages the data are well described by the simulation.

 Fig.\,\ref{fig:Fig-EDrift} compares the measured with the simulated current transients for bias voltages between 100 and 1000\,V for $e$.
 With the exception of the 100\,V data, the data are well described by the simulations.
 The biasing voltage of 100\,V is only 12.5\,V above the depletion voltage, and the effect of the $p^+n$ junction where the charges are generated, and of the $n^+n$ transition, which the electrons reach at the higher drift times, may be significant.

 Fig.\,\ref{fig:Fig-HDrift} compares the measured with the simulated current transients for bias voltages between 100 and 1000\,V for $h$.
 Again, with the exception of the 100\,V data, the data are well described by the simulations.
 At 100\,V the electron-hole pairs are generated by the laser light in the low-field region of the sensor.
 As the density of the generated electron-hole pairs of $5 \cdot 10^{12}$\,cm$^{-3}$ is similar to the doping density, the so called plasma effect\,~\cite{Williams:1974, Becker1:2010} is expected to occur at low bias voltages.
 The plasma effect is due to the shielding of the sensor field by the counter field of the overlapping electron-hole clouds which results in a delayed charge collection.
 In the simulations the plasma effect is not included.
 We note that the measured current transients for voltages between 150 and 1000\,V are well described by the simulations using the transfer function determined by the method described in the paper from the measurement with 1060\,nm laser light at 1000\,V.
 
 The transfer functions have also been determined for the $\langle 100 \rangle $ silicon pad diodes with different thicknesses and capacitances mentioned on page 1. It is found that, although there are significant differences in the transfer functions, the measurements can be well described with consistent values for the field dependence of the electron and hole mobilities. This demonstrates the validity of the proposed method of determining electronic transfer functions.

 \paragraph{Practical aspects of the determination of the transfer function}
 We conclude the manuscript with a few comments on the experience we made, when we developed the method of determining the transfer function described in this Technical Note.
 Initially we used a SPICE simulation of detector, cables and readout electronics.
 Details are given in Ref.\,~\cite{Becker:2010}, where it is also shown that a number of parasitic elements had to be included in the simulation to achieve an acceptable description of the measurements.
 However, the results never have been fully satisfactory especially for diodes with a capacitance above full depletion below 10\,pF.
 We then used the method described here and obtained satisfactory results from the beginning.
 To better understand the method and its possible applications, the following studies have been made.

 As we do not know, if the current recorded by the oscilloscope is the instantaneous current or the current averaged over the sampling interval, the difference between a simulation using the value at the bin center and the average value has been investigated.
 Differences were observed at the maxima and minima of the transients, however, they are smaller than $1.5$\,\% of the maximum signal and do not change significantly the results of the analyses.

 For the Fast-Fourier-Transform of the spline-interpolated measurement $I^{int}$ we included $2$\,ns of the measurement before the pulse has reached $10$\,\% of the maximum. Additionally, we had to add at least 3 bins ($30$\,ps) which are set to zero at the start of $I^{int}$. Otherwise, we sometimes experienced oscillations in the transfer function.

 We have investigated which of our data should be used for obtaining the optimal transfer function.
 We find that the best overall description of the complete data set is obtained, if the $IR$ measurement at 1000\,V is used.
 Our explanation is that the uncertainties of the simulation are smallest for these conditions.
 At the $n^+n$ and  $p^+n$ sides, there are shallow non-depleted regions as well as high-field regions from the doping gradients, which have not been taken into account in the simulation.
 For the red laser a significant fraction of the $eh$\,pairs is generated in these regions, whereas for the infrared laser the charges are generated throughout the sensor.
 In addition, at 1000\,V the electric field in the $200\,\upmu $m thick sensor is high; therefore, the variations of the drift velocities, which approach the saturation velocities, are  minimal.
 We conclude: The precise simulation of the transient induced in electrodes of the sensor is one necessary condition for the successful application of the proposed method.

 Next we discuss the requirements for noise and sampling frequency.
 Ideally the sampling step should be small compared to the width of the transfer function.
 In our example, however, the sampling step of 200\,ps is not much shorter than the measured rise time of about 600\,ps.
 As the simulated rise time of the current induced in the electrodes is much shorter, the measured rise time is approximately equal to the width of the main peak of the transfer function.
 However, if the noise of the transient measurement is small, which has been achieved by injecting $10^6\,eh$\,pairs per pulse and averaging 512 pulses, an interpolation of the measurements can give reliable results.
 We have used a 5th order spline interpolation for obtaining data in 10\,ps steps from the measured transients which were recorded in 200\,ps steps.
 Thus, the second necessary  condition for a successful application of the proposed method are a high signal-to-noise ratio and sampling steps shorter than the rise time of the transient. 

 Last but not least: the transfer function, which has been determined from a measurement for which the intrinsic detector response could be simulated precisely, can only be used for analyzing data from a sensor with similar parameters, with the capacitance being the most important one.

\section*{Acknowledgments}
%\label{sect:Acknowledgements}
 We would like to thank Julian Becker for making available his program which simulates the current transients in silicon pad diodes and for many helpful discussions, J\"orn Schwandt for making the TCAD simulations of the electric field which takes into account the $p^+n$ and $n^+n$ transition regions, and Erika Garutti and Ulrich Koetz, who proof-read the manuscript and made valuable suggestions.
 We are also grateful to the HGF Alliance \emph{Physics at the Terascale} which had funded the TCT set-up used for the measurements.

\section*{Bibliography}
%\label{sect:Bibliography}

\bibliography{science_.bib}
\bibliographystyle{unsrt}

\end{document}